\documentstyle[preprint,aps]{revtex}
\newcommand{\doublespace}{
    \renewcommand{\baselinestretch}{1.6}\large\normalsize}

\newcommand{\bce}{\begin{center}}
\newcommand{\ece}{\end{center}}
\newcommand{\be}{\begin{equation}}
\newcommand{\ee}{\end{equation}}
\newcommand{\bea}{\vspace{0.25cm}\begin{eqnarray}}
\newcommand{\eea}{\end{eqnarray}}

\def\PLA{{Phys. Lett.}  A }
\def\PLB{{Phys. Lett.}  B }
\def\PRL{{Phys. Rev. Lett.} }
\def\PRA{{Phys. Rev.} A }

\def\PRD{{Phys. Rev.} D }

\doublespace

\begin{document}

\title{{\LARGE {\bf Conclusive tests of local realism and pseudoscalar mesons }}}

\doublespace

\author{M.Genovese \footnote{ genovese@ien.it. Tel. 39 011 3919234, fax 39 011 3919259}, C.Novero}
\address{Istituto Elettrotecnico Nazionale Galileo Ferraris, Str. delle Cacce
91,\\I-10135 Torino }
\author{E. Predazzi}
\address{Dip. Fisica Teorica Univ. Torino and INFN, via P. Giuria 1, I-10125 Torino }
\maketitle

\vskip 1cm
{\bf Abstract}
\vskip 0.5cm 
Many beautiful experiments have been addressed to test standard quantum mechanics against local realistic models. Even if a strong evidence favouring standard quantum mechanics is emerged, a conclusive experiment is still lacking, because of low detection efficiencies. Recently, experiments based on pseudoscalar mesons have been proposed as a way for obtaining a conclusive experiment. In this paper, we investigate if this result can effectively be obtained. Our conclusions, based on a careful analysis of the proposed set ups,  are that this will not be possible due to intrinsic limitations of these kind of experiments.   
\vskip 1.5cm


 
\newpage
\bigskip
\centerline{\bf I. INTRODUCTION}
\bigskip
Quantum Mechanics (QM) is one of the pillar of modern physics and no doubt can remain on its validity,  thanks to the huge amount of experimental results  confirming many its predictions (often to an extremely precise level).
Nevertheless, some doubt could persist if a general local realistic theory could include QM as a statistical limit.

This idea appeared already in 1935 in the famous Einstein-Podolsky-Rosen paper \cite{EPR}.  

For this purpose, the authors introduced the concept of { \it element of reality} according to the following
definition: If, without disturbing in any way a system, I can predict
without any uncertainty the value of a physical quantity, then there is an element of physical reality corresponding to this quantity. They formulated also the reasonable hypothesis (in the light of special relativity) that any non-local action was forbidden.

They concluded  that either one of their hypothesis was wrong or
Quantum \ Mechanics was not a complete theory, in the sense that not every
element of physical reality had a counterpart in the theory.

With the purpose of elucidating their arguments, we will  follow their discussion, but with the physical system introduced by
Bohm, which allows a clearer comprehension of the problem.

Let us thus consider a singlet state of two spin 1/2 particles

\begin{equation}
\vert \psi \rangle = {\frac{ \vert 1 \rangle \vert -1 \rangle - \vert -1
\rangle \vert 1 \rangle }{\sqrt {2}} }
 \label{Singlet}
\end{equation}
where $\vert \pm 1 \rangle $ represents a single particle of spin up and down
states, respectively. This is an example of an entangled state, namely of a state of two or more particles which cannot be factorised in single-particle states.

Let the two particles  separate and then measure the z
component of the spin of the first particle; this permits us to know
immediately the z spin component of the particle 2 (which is opposite) without disturbing in any way the second particle. Thus the z component of the spin of the second particle is an element of reality according to the previous
definition.

But, since the singlet state is invariant under rotations, we could
refer to any other axis (as x or y etc.): thus we can argue that any other
spin component of particle 2 is an element of reality. However, spin components on different axis are incompatible variables in Standard Quantum Mechanics (SQM), for which one cannot assign definite values at the same time: for this reason it follows that Quantum Mechanics cannot be a complete theory, for it does not allow a prediction of all elements of reality.

This statement was somehow the starting point of development of the so-called
Local Realistic Models (LRM or hidden variables theory):  it exists a
deterministic (and local) theory  describing nature and  Quantum
Mechanics is only a statistical approximation of this theory. 

In particular in any hidden variable theory every particle has a perfectly
assigned value for each observable, determined by a hidden variable $x$. A
statistical ensemble of particles has a certain distribution $\rho (x)$
of the hidden variable and thus the average value of an observable A is
given by:

\bigskip

\begin{equation}
<A>=\int dx\rho (x)A(x)  \label{aa}
\end{equation}

Of course, considering the great success of SQM in predicting many different
experimental data, the average $<A>$ given by (\ref{aa}) must reproduce the
SQM predictions.

Historically, the quest for hidden variable theories  stopped
when von Neumann published a theorem asserting the
impossibility of constructing a hidden variable theory reproducing all the
results of SQM. For long time, his prestige led to an acritical acceptation of this theorem, but it was then discovered that one of his hypotheses was too restrictive so that the program of a hidden variable theory was still possible.

The fundamental progress in discussing possible extensions of SQM was  Bell's proof \cite{Bell} that any realistic Local Hidden Variable LRM theory must satisfy certain inequalities which can be violated in QM. This allows, in principle, testing  experimentally  the validity of SQM against LRM.

Various different forms of these inequalities have been proposed. All of them are equivalent by the Fine theorem (see e.g. Ref. \cite{A}).  
For the sake of exemplification let us show one of them. 
Let us consider a source of entangled particles emitting a pair, where the first particle goes to detector 1 and the second to detector 2, and suppose that before the detector i we select a certain property $\theta _{i}$. For example, if the particles are entangled in spin, then $\theta _{i}$  is the angle defining the direction (respect to the z-axix) along which we are going to measure the spin. The Clauser-Horne sum then reads:

\bigskip

\begin{equation}
CHS=P(\theta _{1},\theta _{2})-P(\theta _{1},\theta _{2}^{\prime })+P(\theta
_{1}^{\prime },\theta _{2})+P(\theta _{1}^{\prime },\theta _{2}^{\prime
})-P(\theta _{1}^{\prime })-P(\theta _{2})  \label{eq:CH}
\end{equation}
where $P(\theta _{1},\theta _{2})$ \ represents the joint probability of
observing a particle in 1 with the selection $\theta _{1}$ and, in
coincidence, \ a particle in 2 with the selection $\theta _{2.}$
On the other hand, $P(\theta _{i})$ \ represents the probability of
observing a single particle at i with selection $\theta _{i}$. 

Clauser and Horne \cite{CHS} have shown that CHS  is always smaller than zero for every LRM, but it can be larger than zero for a suitable choice of parameters in SQM.

Many interesting experiments have been performed to  test  Bell inequalities. Most of them were  based on entangled photon pairs produced in positronium decays, atomic decays and in non-linear crystals.  Only few exceptions  were performed with other systems \cite{pp,Win}. All of them led to a substantial agreement with quantum mechanics \CITE{CH,Mandel,asp,franson,type1,type2}, bar one \cite{catania} not confirmed by similar ones \cite{cat2}. Thus LRM theories are disfavoured (see ref.s \cite{rev} for reviews), but, so far, no experiment has yet been able to exclude definitively such theories. In fact, so far,  one has always been forced to introduce a further additional hypothesis \CITE{santos}, due to the low total detection efficiency, stating that the observed sample of particle pairs is a faithful subsample of the whole. This problem is known as { \it detection or efficiency loophole}. For example, in Eq. \ref{eq:CH} this reflects into the fact that, because of losses, single particle probabilities $P(\theta _{i})$ are always much larger than joint probabilities 
$P(\theta _{i},\theta _{j})$ and thus in real experiments must be substituted with joint probabilities without selection on one arm.
The research for new experimental configurations able to overcome the detection loophole is of course of the greatest interest. 

Since the beginning of 90's, big progresses in this direction have been obtained by using parametric down conversion (PDC) processes for generating entangled photon pairs with an high angular correlation. 

The generation of entangled states by parametric down conversion (PDC) has replaced other techniques for it overcomes some former limitations. In particular, having angular correlations better than 1 mrad, it overcomes the poor angular correlation of atomic cascade photons, as it was in the celebrated experiment of A. Aspect et al. \CITE{asp}.  This was at the origin of the small total efficiency of this type of experiments in which one is forced to select a small subsample of the produced photons, leading inevitably to  the detection loophole. Furthermore, a good polarisation selection can be done using visible light, at variance with the case of positronium decay, where this was not possible for high energy gamma rays (in spite of a very high detection efficiency). 

The first experiments using this technique were performed with type I PDC, which gives phase and momentum entanglement and can be used for  testing  Bell inequalities using two spatially separated interferometers \cite{franson}, as realised by \cite{type1}. The use of beam splitters, however, strongly reduces the total quantum efficiency. 

Alternatively, one can generate a polarisation entangled state \cite{ou}. However,  the first experiments for the creation of couples of photons entangled from the point of view of polarisation, still suffered severe limitations, as it was pointed out in Ref. \cite{santos}. The essence of the problem is that in generating this state, half of the initial photon flux was lost, and one was, of necessity, led to assume that the photons' population actually involved in the experiment to be a faithful sample of the original one, i.e. falling on the efficiency loophole. 

More recently, experiments  have been realised using Type II PDC \cite{type2}, where a polarisation entangled state is directly generated. This scheme has permitted, at the price of delicate compensations for having identical arrival time of the ordinary and extraordinary photon, a much higher total efficiency than the previous ones (up to about $0.3$), even though this is still far from the required value of $0.81$. In addition, some recent experiments studying equalities among correlations functions of two \cite{dem} or three photons \cite{GHZ} rather than Bell inequalities are far from solving these problems \cite{garuccio}. A large interest remains therefore for new experiments increasing total quantum efficiency in order to reduce and finally overcome the efficiency loophole. 

Some years ago, a very important theoretical step in this direction has been performed recognising that, while for maximally entangled pairs a total efficiency larger than to 0.81 is required to obtain an efficiency-loophole free experiment, for non maximally entangled pairs this limit is reduced to 0.67 \cite{eb} (in the case of no background). However, it must be noticed that, for non-maximally entangled states, the largest discrepancy between quantum mechanics and local hidden variable theories is reduced: thus a compromise between a lower total efficiency and a still sufficiently large value of this difference will be necessary when realising an experiment addressed to overcome the detection loophole.

 An experiment devoted to test Bell inequalities using non-maximally entangled photon pairs has been recently realised \cite{nos}. In this scheme the non-maximally entangled states of photons in polarisation were obtained superimposing through an optical condenser the parametric down conversion produced by two type I crystals, separated of a distance much smaller than the coherence one of the pump laser. This scheme allows to obtain a very good superposition, not possible in the scheme of Ref. \cite{Kwn}, and a good spatial separation of the two substates, not possible with the very recent scheme of Ref. \cite{Shn}.

Even if relevant progresses toward the elimination of the detection loophole have been obtained using entangled photon pairs, nevertheless the total efficiency is strongly dominated by the quantum efficiency of photodetectors.
Nowadays efficiencies for commercial photodetectors are under 75 \%. Prototypes already reach much higher efficiencies \cite{Yam}, but at the prize of an high background which also limits the possibility of a loophole free test \cite{Tho}. 

Thus, in summary, the use of entangled photon pairs has led to very important tests of Bell inequalities, but at the moment cannot allow to eliminate   the detection loophole.

On the other hand, a recent experiment \cite{Win} performed using Be ions has reached very high efficiencies (around 98 \%), but in this case the two subsystems (the two ions) are not really separated systems and the test cannot be considered a real implementation of a detection loophole free test of Bell inequalities \cite{Vai}, even if it constitutes a relevant progress in this sense.  

\bigskip
\centerline{\bf II. BELL INEQUALITIES AND KAONS }
\bigskip

Even if little doubts remain on the validity of the standard quantum mechanics, considering the fundamental importance of the question, the quest for other experimental schemes for a definitive test of Bell inequalities is therefore of the largest interest. 

Many theoretical investigations have been recently devoted to find new schemes eventually suitable for a loophole free experiment \cite{newP}.
In particular, in the last years, a considerable effort has been addressed to investigate the possibility that the use of entangled pseudoscalar meson pairs as $K \bar{K}$ or $B \bar{B}$ ($D$ mesons decay rapidly through Cabibbo allowed channels and therefore do not represent an interesting situation) might lead to a conclusive test of local realism. Effectively, this solution presents some evident advantages: if the pair is produced by the decay of a particle at rest in the laboratory frame (as the $\phi$ at Daphne), the two particles can be easily separated at a relatively large distance allowing an easy space-like separation of the two subsystems and permitting an easy elimination of the space-like loophole, i.e. realising two completely space-like separated measurements on the two subsystems. Let us notice that for a complete elimination of this loophole,  the space-like separation must include the setting of the experimental selections as well (for photons this could be performed with large separations and/or acousto-optical devices for polarisation selection, but introducing big losses which make impossible a contemporary elimination of the detection loophole). Moreover, in these experiments based on pseudoscalar pairs a very low noise is expected.

These proposals are based on the use of entangled states of the form (analogous to the singlet state of Eq. \ref{Singlet}):
\bea
|\Psi \rangle = { | K_0 \rangle | \bar K_0 \rangle  - | \bar K_0 \rangle | K_0 \rangle \over \sqrt{2} } = & \cr
= { | K_L \rangle |  K_S \rangle  - |  K_S \rangle | K_L \rangle \over \sqrt{2} } & \cr
\label{psi}
\eea

where $| K_0 \rangle$ and $| \bar K_0 \rangle$ are the particle and antiparticle related by charge conjugation and composed by a quark $d$ with an anti-strange $\bar s$ and a $\bar d $ with a $s$ respectively. Whilst the CP (charge conjugation and parity) eigenstates 
\be
| K_L \rangle = {| K_0 \rangle  - | \bar K_0 \rangle 
\over \sqrt{2}}
\ee
and
\be
| K_S \rangle = {| K_0 \rangle  + |  \bar K_0 \rangle 
\over \sqrt{2} }
\ee 
are the long living (CP=-1, forbidding the 2 pions decay) and the short living (CP=+1, allowing the 2 pions decay) states. Here and in the following small effects due to CP violation in weak interactions are neglected, not being relevant for our discussion.
 
The high efficiency of particles detectors has led to the claim that  tests of local realism realised with pseudoscalar mesons pairs could allow the elimination of the detection loophole. The purpose of this article is a critical analysis of this statement. 

First of all it must be noticed that a simple test based on a correlation function defined such that it takes the value 1 when two or none $\bar K_0$ are identified and -1 otherwise, would not lead to a violation of Bell inequalities due to the specific values of $ K_0 \bar K_0 $ mixing parameters, as shown in Ref. \cite{gh}. Also, a simple hypothesis \cite{furry} where the state  (\ref{psi}) collapses shortly after its production in two factorised states
\be
|  K_S \rangle | K_L \rangle , \, |  K_L \rangle | K_S \rangle 
\ee
is already excluded \cite{BF}.
Anyway, other Bell inequalities and hidden variable schemes can be considered \cite{BellK,DG,Sel1,Sel2}.

Our main concern about the statement that detection efficiency is high for mesons comes from the fact that in any experimental test proposed up to now one must tag the $P$ or $\bar{P}$ through its decay. This requires the selection of $\Delta S = \Delta Q$ semileptonic decays, which represent only a fraction of the total possible decays of the meson.
From Ref. \cite{PDB} one derives the following branching ratios:
\bea
 BR(K^0_S \rightarrow \pi^{+} e^{-} \nu_e) = (3.6 \pm 0.7) 10^{-4} \cr
BR(K^0_L \rightarrow \pi^{+} e^{-} \nu_e) = 0.1939 \pm 0.0014  \cr
BR(K^0_L \rightarrow \pi^{+} \mu^{-} \nu_{\mu}) = 0.1359 \pm 0.0013 \cr
BR(B^0 \rightarrow  l^{+} \nu_{l} X) = 0.105 \pm 0.008 \cr
BR(B^0 \rightarrow  l^{+} \nu_{l} \rho ^-) = (2.6 \pm 0.7) 10^{-4} \cr
BR(B^0 \rightarrow  l^{+} \nu_{l} \pi ^-) = (1.8 \pm 0.6) 10^{-4}
\label{BR}
\eea
where $X$ means anything.

Besides this problem, one has to consider experimental cuts on the energies of the decay products, which  inevitably further reduce this fraction. Moreover, a fraction of the pairs will be lost by decays occurring before the region of observation. Finally, most of these proposals involve the regeneration phenomenon, which introduces further strong losses; on the other hand if no choice of the measurement set up (e.g. the presence of the regeneration slab) is introduced, the space-like loophole cannot be really eliminated. 

The result of these considerations is that one is unavoidably led to subselect a fraction of the total events. As one cannot exclude a priori  hidden variables related to the decay properties of the meson and losses, one cannot exclude the sample to be biased and thus the detection loophole appears here too.
This is analogous to the photons experiments, where the detection loophole derives by the fact that one cannot exclude  losses  related to the values of hidden variables which determine if the photon passes or not a polarisation (or another) selection.  Namely, in a local realistic model the properties of a particle are completely specified by the hidden variables. Also decays, in a deterministic model, can happen according to the values of the hidden variables (both in a deterministic or in a probabilistic way). Thus, states with different hidden variables can decay in different channels, with the condition that the branching ratios {\it averaged } over the hidden variables distribution reproduce the quantum mechanics predictions. Analogously, any other loss can depend on hidden variables as well.

For what concerns the experiments based on Bell inequalities \cite{BellK,DG}, the same limits for the total efficiency previously discussed remain valid. As the total branching ratio in  $\Delta S = \Delta Q$ semileptonic decays is much smaller than 0.81, this unavoidably implies that a loophole free test of Bell inequalities cannot be  performed in this class of experiments. The eventual use of non-maximally entangled states,  lowering the efficiency threshold to 0.67, does not change the situation. This problem does not appear in Ref. \cite{BF}, however other additional hypotheses are needed (see Eq. 15 and discussion after Eq. 18 of \cite{BF}), and thus this proposal does not allow a general test of LRM as well.

It must also be noticed  that the only observation of interference between the two terms of the entangled wave function, Eq. \ref{psi}, as in Ref \cite{CLEO}, does not exclude  general LRM, for this feature can be reproduced in a general class of local realistic theories. 

In summary, we conclude that Bell inequalities measurement on 
kaons pairs will not allow a definitive test of local realism.
 
\bigskip
\centerline{\bf III. OTHER TESTS OF LOCAL REALISM WITH KAONS}
\bigskip

Let us then consider other proposals for testing local realism, not based on a  Bell inequalities measurement. Two proposals of this kind have been recently advanced by F. Selleri and others concerning a  $K \bar{K}$ \cite{Sel1} (very lately further developed in Ref. \cite{DG}) or  a $B \bar{B}$ \cite{Sel2} system respectively.

In order to show how the problem previously discussed appears for these proposals as well, let us analyse in details the $K \bar{K}$ case as discussed in Ref. \cite{Sel1}. The $B \bar{B}$ one will follow with small modifications.

In the model of Ref. \cite{Sel1}, the $K \bar{K}$ pair is local-realistically described by means of two hidden variables, one ($\lambda_1$) determining a well defined CP value, the other ($\lambda_2$) a well defined strangeness $S$ value for the $K$ (and related to this for the $\bar K$). This second variable cannot be a time independent property, but is subject to sudden jumps. If locality must be preserved the time of this jump must already be fixed ab initio by a hidden variable $\lambda _3$ (which represents the real second hidden variable of the model) and the two subsystems must not influence each other while they are flying apart, namely $\lambda_2$ is not the true hidden variable, but a parameter driven by $\lambda _3$ (see appendix of Ref. \cite{Sel1} and Ref. \cite{DG}).

Following Ref. \cite{Sel1}, we  denote by $K_1$ the state with CP=1, S=1, $K_2$ the state with CP=1, S=-1,
$K_3$ the state with CP=-1, S=1 and
$K_4$ the state with CP=-1, S=-1.

The initial state can be, with probability $1/4$, in anyone of the states $CP= \pm 1$, $S=\pm 1$. Each of these pairs give, in the local-realistic model (LRM), a certain probability of observing a $\bar K_0 \bar K_0$ pair at proper times $t_a$ and $t_b$ ($\ne t_a$) of the two particles. These probabilities are (in a somehow simplified form, see eq. 62-70 of Ref. \cite{Sel1}):
\bea
P_1[t_a,t_b]=[E_S(t_a) Q_-(t_a) - \rho(t_a)] \cdot E_L(t_a) p_{43}(t_b | t_a) & \cr
P_2[t_a,t_b]=[E_S(t_a) Q_+(t_a) + \rho(t_a)] \cdot E_L(t_a) p_{43}(t_b | t_a) & \cr
P_3[t_a,t_b]=[E_L(t_a) Q_-(t_a) + \rho(t_a)] \cdot E_S(t_a) p_{21}(t_b | t_a) & \cr
P_4[t_a,t_b]=[E_L(t_a) Q_+(t_a) - \rho(t_a)] \cdot E_S(t_a) p_{21}(t_b | t_a) \, \,
\label{P} 
\eea
corresponding to an initial state with $K_1$ on the left and $K_4$ on the right,
$K_2$ on the left and $K_3$ on the right, $K_3$ on the left and $K_2$ on the right and $K_4$ on the left and $K_1$ on the right respectively.

In Eq. \ref{P}, we have introduced $ E_S(t) = exp(- \gamma_S t)$ and $ E_L(t) = exp(- \gamma_L t)$, where (in units $c=\hbar=1$) $\gamma_{S}=(1.1192\pm 0.0010) 10^{10} s^{-1}$
and $\gamma_{L}=(1.934 \pm 0.015) 10^{7} s^{-1}$ denote the decay rate of $K_S$ and $K_L$ \cite{PDB}. 

Furthermore, the function $Q_{\pm}$ are defined through:
\be
Q_{\pm} ={ 1 \over 2}  \left[ 1 \pm {2 \sqrt{E_L E_S} \over E_L + E_S} \cos( \Delta m t) \right]
\ee
where $\Delta m = (0.5300 \pm 0.0012) 10^{10} s^{-1}$ is the mass difference $ M_{K_L} -M_{K_S}$.
We have also introduced the symbol $p_{ij}(t_a | t_b)$ for denoting the probability of having a $K_i$ at time $t_b$ conditioned to have had the state $K_i$ at time $t_a$. From Ref. \cite{Sel1} one has: 
\be
 p_{21}(t_b | t_a) = E_S^{-1}(t_a) [p_{21}(t_b | 0) - p_{21}( t_a | 0)  \cdot E_S(t_b-t_a)]
\ee
and
\be
p_{43}(t_b | t_a) = E_L^{-1}(t_a) [p_{43}(t_b | 0) - p_{43}( t_a | 0)  \cdot E_L(t_b-t_a)]
\ee

where
\be
 p_{21}(t | 0) = E_S(t) Q_-(t) - \rho(t)
\ee 
and 
\be
 p_{43}(t | 0) = E_L(t) Q_-(t) + \rho(t).
\ee
Finally, $\rho(t)$ is a function not perfectly determined in the model (see discussion in Ref. \cite{Sel1}), but which is limited  by
\bea
-E_S Q_+ \le \rho \le E_S Q_- \cr
-E_L Q_- \le \rho \le E_L Q_+ \cr
\label{ro}
\eea

If we had  a total efficiency of 1, the LRM probability of observing a $\bar K_0 \bar K_0$ pair is given by the sum of the four probabilities of Eq. \ref{P} multiplied by $1/4$. It is rather different from the quantum mechanical prediction 
\be
P_{QM}[\bar{K_0}(t_a), \bar{K_0}(t_b)]= {1 \over 8} \left [ e^{[-(\gamma_S t_a + \gamma _L t_b)]} +
 e^{[-(\gamma_L t_a + \gamma _S t_b)]} - 2 e^{[-(1/2) (\gamma_S + \gamma_L) (t_a + t_b)] } \cos (\Delta m (t_a - t_b)) \right]
\ee 
and thus represents a good test of the LRM. This is shown in fig. 1, where  $ P[\bar{K_0}(t_a), \bar{K_0}(2 t_a)]$ is reported in analogy with Table 1 of Ref. \cite{Sel1}.

However, when the total efficiency is lower than 1, the different probabilities can contribute in different ways as the hidden variables, which determine the passing or not the test, could  also be related to the decay properties of the meson pair and losses.  As discussed previously, the specific property of the meson is not being or not a $\bar K_0$ at a certain proper time, but the hidden variables values characterise it completely, and thus, in principle, even its decay properties. 
If this is the case, different coefficients $a_i$   can multiply the four probabilities. One has therefore:
\be
P[\bar{K_0}(t_a), \bar{K_0}(t_b)]= 1 / 4 \cdot \left[ a_1 P_1 [t_a,t_b] + a_2 P_2 [t_a,t_b] + a_3 P_3 [t_a,t_b] + a_4 P_4 [t_a,t_b] \right]
\ee

The freedom of the choice of these parameters allows to reproduce the quantum mechanical prediction.
In figure 2, we show how the lower limit of the LRM can be smaller than the SQM curve, when this effect is considered for the curves of fig.1. We have chosen the parameters set: $a_1=1$,$a_2= 0.07$,$a_3= 0.03$ and $a_4=0.1$. This corresponds to values of $a_i$  such that their sum is the total efficiency chosen to be 0.3 (other values can be obtained by scaling). Nevertheless, it must be noticed that the values of the $a_i$ are substantially very little constrained. Besides, they can depend on the times $t_a$ and $t_b$, like the value of the hidden variable $\lambda_2$: this property, obviously, makes the quantum mechanics prediction more reproducible. Let us also notice that situation 1 and 4, 2 and 3 are symmetric under the exchange of left and right, although the decay probabilities do not need to be the same. 
The only real constraint on $a_i$ is that the total fraction of observed decays should be reproduced. 
For example, a trivial solution which reproduces SQM predictions is  $a_i(t_a,t_b) = P_{QM}(t_a,t_b)/P_i(t_a,t_b)$. Of course, the same discussion can be exactly enforced to other correlation probabilities as $ P[\bar{K_0}(t_a), {K_0}(t_b)]$ etc.

The previous results are obtained with  $\rho=E_s Q_-$, similarly to fig.1, in order to allow a direct comparison with the results of Ref. \cite{Sel1}. 
However, according to the discussion of Ref. \cite{DG}, the value $\rho=0$ is the favoured. We present the results for the case $\rho=0$ in figure 3. A total efficiency of 0.3 is considered. In this case the two curves (SQM and LRM) are very similar, for other choices of the parameters the lowest limit of the LRM could be easily reduced further. It may be interesting to notice that similar results are obtained also with other choices of $\rho[t]$, compatible with Eq. \ref{ro}. 

Altogether, these results show that  the lowest limit curve of local realism can easily reproduce or be lower than the quantum mechanics prediction when different weights multiply the four probabilities, due to different branching ratios and losses for the 4 cases. It must be emphasised that our purpose is only to show a counterexample, which proves that observation of the curve predicted by SQM cannot exclude {\it every } LRM. Our results show unequivocally that such a counterexample exists.

It must also be noticed that if not only the decay (and losses) properties were related to the hidden variables, but also the decay time were determined by them, then the scheme we have discussed should be substantially modified leaving a even larger space for agreement between SQM and LRM. 
 
Of course, as all LRM tests performed up to now, a result in agreement with SQM,  will further reduce the space for the existence of a LRM, even if unable to obtain the general result of eliminating the possible existence of LRM.
  
In conclusion, when only a subsample is selected  (semileptonic  $\Delta S = \Delta Q$ decays must be observed for tagging a $\bar K_0 \bar K_0$ state and cuts must be introduced) the result of this analysis shows that detection loophole appears also in this case. Of course we are not discussing how the local realistic  model should be, or if its complexity makes it unpleasant, we are simply investigating if such experiments allow to exclude without any doubt every local realistic model: our conclusion  is that one is obliged to introducing the additional hypothesis that the observed sample is unbiased concerning the hidden variables values, i.e. the efficiency loophole reapers.

\bigskip
\centerline{\bf IV. TESTS OF LOCAL REALISM WITH $B_0$}
\bigskip

 Precisely the same considerations can be carried out for  entangled $B_0 \bar{B_0}$ pairs (produced, for example, in $\Upsilon (4S)$ decays). 

A first set of papers \cite{BF} consider the possibility to test SQM measuring the term deriving from interference between the two terms in the entangled wave function (\ref{psi}). However, this effect is also reproduced in any reasonable LRM \cite{Sel2} and thus cannot be considered a general test of local realism, but can only allow to eliminate some specific class of hidden variable theories.

For what concerns Bell inequalities, exactly the same discussion of paragraph II applies: also in this case one will have to tag the $\bar B_0$ by semileptonic decays (see the branching ratios of Eq. \ref{BR}) and the detection efficiency is expected to be around $45 \%$ \cite{Beff}. Therefore we conclude that for $B$ pairs as well no definitive test of Bell inequalities is possible.

The same conclusions are reached for tests of local realism without inequalities, for the scheme of Ref. \cite{Sel2} is equivalent to the one we have discussed in the previous paragraph,  a part small changes due to the same decay width $\gamma = (0.646 \pm 0.013) 10^{12} s^{-1}$ \cite{PDB} for both the CP eigenstates. 
More in details the SQM prediction simplifies into:
\be
P_{QM}[\bar{B_0}(t_a), \bar{B_0}(t_b)]= {1 \over 4} \left [ e^{[-\gamma (t_a +  t_b)]} [1 - \cos (\Delta m (t_a - t_b))] \right]
\ee

Also the probabilities $P_i$ simplifies analogously. For example the functions $Q_{\pm}$ become:
\be
Q_{\pm}[t] = { 1 \over 2} [ 1 \pm \cos( \Delta m t)]
\ee
where $\Delta m = ( 0.472 \pm 0.017 ) 10^{12} s^{-1}$ \cite{PDB}.

The SQM prediction and the LRM ($\rho = 0$) one are compared in fig.4 in analogy to fig.3 for the kaon case, both for the cases of detection efficiency equal to one and when the discussed effect is introduced.
The figure exemplifies as the same results of the $K \bar K$ case keep their validity: in presence of a detection efficiency smaller than 1, LRM results can simulate SQM ones. Of course, as for some LRM the probabilities $ P[\bar{B_0}(t_a), \bar{B_0}(t_b)]$ and $ P[{B_0}(t_a), \bar{B_0}(t_b)]$ can reproduce  the SQM result, also the asymmetry 
\be
A[t_a,t_b]= { P[\bar{B_0}(t_a), \bar{B_0}(t_b)] - P[{B_0}(t_a), \bar{B_0}(t_b)] \over P[\bar{B_0}(t_a), \bar{B_0}(t_b)] + P[{B_0}(t_a), \bar{B_0}(t_b)] }
\ee
discussed in Ref. \cite{Sel2},  and the ratio 
\be
R = { P[\bar{B_0}, \bar{B_0}] + P[{B_0}, {B_0}] \over P[\bar{B_0}, {B_0}] + P[{B_0}, \bar{B_0}] }
\ee
suggested in Ref. \cite{Home} (where  time integrated probabilities appear),
could be equal to the SQM ones for some LRM, and therefore cannot be used for a conclusive test of local realism.

Altogether, our conclusion is therefore that  the $ B_0 \bar B_0$ case does not allow an experiment free of the detection loophole  as well. 

\bigskip
\centerline{\bf V. CONCLUSIONS}
\bigskip

In summary, the result of this analysis shows that the detection loophole appears in a unavoidable way in experiments addressed to test local realism by the use of pseudoscalar mesons entangled pairs.
 
This is due to the fact that losses and decay properties might be related to the value of the hidden variables and therefore the selected subsample of the initial pairs might be biased, leading to the so called detection loophole. 

This is a general problem, which is unavoidable every time the particle must be identified by using a particular decay channel (which does not have a branching ratio above 81 \%). This is the case for $K_0$ or $B_0$ tagging, where semileptonic channels are considered. 

 Therefore, even if these experiments represent an interesting new approach for investigating quantum non-locality in a new sector, nevertheless they cannot lead to a conclusive test of local realism, for the impossibility of eliminating the detection loophole.
 
Also other tests of local realism in particle physics would be affected by the same problem, if based on unstable particles which must be tagged using some specific decay channel. On the other hand, this problem would not appear if one  considers tests based on entangled pairs of stable particles, which can be identified without strong losses, as, for example, a singlet pair of $e^+ e^-$ produced in the decay of a scalar particle.  
Anyway, these tests must be performed \cite{ADD} using measurement of non-commuting variables (as different components of spin) and not only a commuting set (as in Ref.s \cite{pp,Tor}) and thus a practical realisation will be far from being trivial.

\bigskip
\centerline{\bf Acknowledgements}
\bigskip
\noindent We would like to acknowledge support of ASI under contract LONO 500172, of  MURST via special programs "giovani ricercatori" Dip. Fisica Teorica Univ. Torino and of Istituto Nazionale di Fisica Nucleare.

\vfill \eject

\bigskip
{\bf Figures Captions}

Fig.1   The SQM (thick) and the minimal  LRM (dashed) predictions for 
$ P[\bar{K_0}(t_a), \bar{K_0}(2 t_a)]$. The result is shown for $\rho[t]= E_S[t] Q_-[t]$, as in table I of Ref. \cite{Sel1}. The minimal LRM  is largely above the SQM prediction. 

Fig.2  The SQM (thick) and the minimal LRM (dashed) predictions for $ P[\bar{K_0}(t_a), \bar{K_0}(2 t_a)]$, as in fig.1 but keeping into account a total detection efficiency of 0.3. With use of the choice $a_1=0.5, a_2=0.13, a_3=0.5, a_4= 0.07$. 

Fig.3  The SQM (thick, labelled with A) and the minimal LRM, $\rho=0$, (dot-dashed, B) predictions for $ P[\bar{K_0}(t_a), \bar{K_0}(2 t_a)]$ keeping into account a total detection efficiency of 0.3. With the choice $a_1=1, a_2=0.13, a_3=0.03, a_4= 0.04$ the two curves substantially coincides. For other choices of the parameters the lower bound of LRM is well under SQM prediction. The four probabilities $P_i[t_a,2 t_a]$ are shown for the sake of completeness, with dashing decreasing in order of the suffix $i$ and labelled by C,D,E,F respectively.

Fig.4  The SQM (thick) and the minimal LRM, $\rho= 0$, (dashed) predictions for $ P[\bar{B_0}(t_a), \bar{B_0}(2 t_a)]$, without (upper curves) and with (lower curves) keeping into account a total detection efficiency of 0.3  (with the choice $a_1=0.52, a_2=0.08, a_3=0.52, a_4= 0.08$).

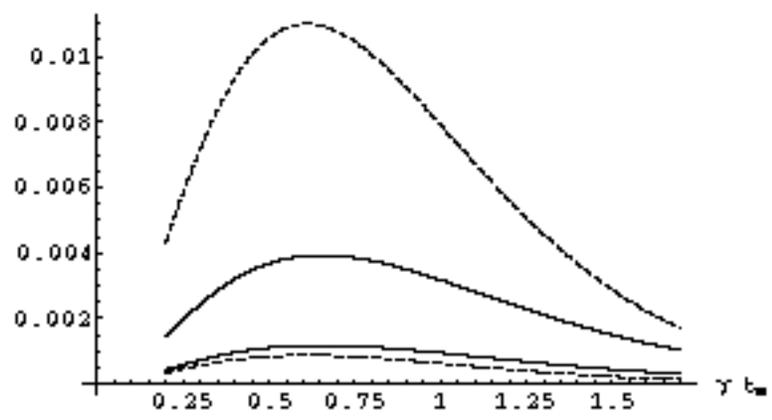
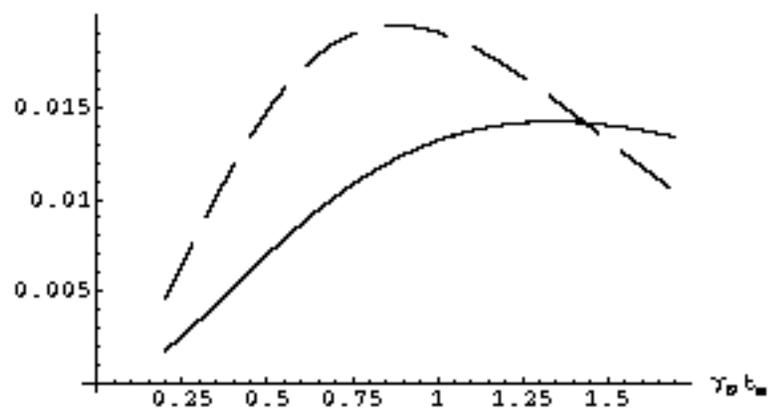
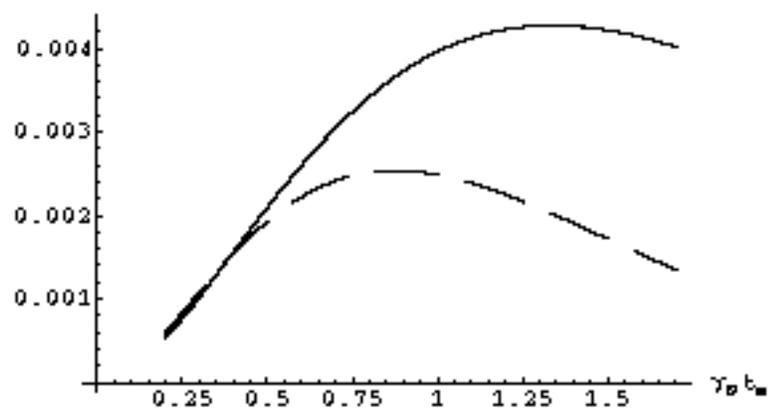
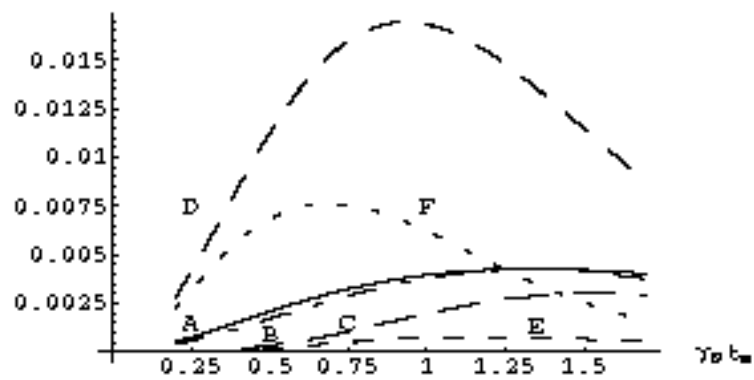
\end{document}